\documentclass[12pt]{article}

\newcommand\eq[1] {(\ref{#1})}

\newcommand\labfig[1] {\label{fig:#1}}

\newcommand\proof{{\it Proof:}\quad}

\newcommand\qed{\hfill $\rlap{$\sqcap$}\sqcup$}

\newcommand{\beqa}{\begin{eqnarray}}
\newcommand{\eeqa}[1]{\label{#1}\end{eqnarray}}
\newcommand{\beq}{\begin{equation}}
\newcommand{\eeq}[1]{\label{#1}\end{equation}}
\newcommand{\ee}{\end{equation}}

\newcommand{\RE}{{\rm I\! R}}

\newcommand{\Gk}{\kappa}

\newcommand{\Go}{\omega}

\newcommand{\GD}{\Delta}

\newcommand{\CA}{{\cal A}}

\newcommand{\CU}{{\cal U}}

\newcommand{\CW}{{\cal W}}

\def\Ba{{\bf a}}
\def\Bb{{\bf b}}

\def\Bn{{\bf n}}

\def\Bu{{\bf u}}
\def\Bv{{\bf v}}
\def\Bw{{\bf w}}
\def\Bx{{\bf x}}

\def\BA{{\bf A}}

\def\BF{{\bf F}}

\def\BI{{\bf I}}

\def\BM{{\bf M}}

\def\BS{{\bf S}}

\def\BV{{\bf V}}
\def\BW{{\bf W}}

\def \ba {\begin{array}}
\def \ea {\end{array}}
\newtheorem {Thm} {Theorem}
\newtheorem {Arem} {Remark}
\newtheorem {Aexa} {Example}
\def \refe #1.{(\ref{#1})}
\def \reff #1.{figure~\ref{#1}}
\def \refs #1.{section~\ref{#1}}
\def \refss #1.{subsection~\ref{#1}}
\def \refD #1.{Definition~\ref{#1}}
\def \refT #1.{Theorem~\ref{#1}}
\def \refL #1.{Lemma~\ref{#1}}
\def \refC #1.{Corollary~\ref{#1}}
\def \refP #1.{Proposition~\ref{#1}}
\def \refR #1.{Remark~\ref{#1}}
\def \refE #1.{Example~\ref{#1}}
\def \refN #1.{Notation~\ref{#1}}

\usepackage{graphics}
\usepackage{amssymb}
\begin{document}
\vspace{-1in}
\title{Realizable response matrices of multiterminal electrical, acoustic, and elastodynamic networks
at a given frequency}
\author{Graeme W. Milton\\
\small{Department of Mathematics, University of Utah, Salt Lake City UT 84112, USA}\\
Pierre Seppecher\\ 
\small{Laboratoire d'Analysis Non Lin{\'e}aire Appliqu{\'e}e et Mod{\'e}lisation},\\
\small{Universit{\'e} de Toulon et du Var, BP 132-83957 La Garde Cedex, France}}
\date{}
\maketitle
\begin{abstract}
We give a complete characterization of the possible response matrices 
at a fixed frequency of
$n$-terminal electrical networks of inductors, capacitors, resistors and grounds,
and of $n$-terminal discrete linear elastodynamic networks of springs and point
masses, both in the three-dimensional case and in the two-dimensional case.
Specifically we construct networks which realize {\it any} response matrix which
is compatible with the known symmetry properties and thermodynamic constraints
of response matrices. Due to a mathematical equivalence we also obtain a characterization
of the response matrices of discrete acoustic networks.

\end{abstract}
\vskip2mm

\section{Introduction}
\setcounter{equation}{0}
It is well known that composites built from high contrast constituents can have moduli or combinations of 
moduli which are not usually seen in nature. For example, by combining stiff and compliant phases
one can obtain composites with a negative Poisson's ratio, having a high shear modulus
but low bulk modulus \cite{Lakes:1987:FSN, Milton:1992:CMP}.
More generally one can construct anisotropic composites having any desired
positive definite elasticity tensor \cite{Milton:1995:WET, Milton:2002:TOC, Camar:2003:DCS}.
Composites have recently been constructed with a negative
refractive index, having a negative electrical permittivity and a negative magnetic permeability
over some frequency range \cite{Shelby:2001:EVN}. They have also been constructed with a negative effective density
and with a negative effective stiffness \cite{Lakes:2001:EDC, Fang:2006:UMN}
over a range of frequencies. Less well known, though
perhaps more interesting, is the fact that the equations describing the macroscopic behavior of 
composites built from high contrast constituents can be entirely different from those seen
in nature. For example one can obtain  materials with macroscopic non-Ohmic, possibly non-local,  
conducting behavior, even though they conform to Ohm's law at the microscale \cite{Khruslov:1978:ABS, Briane:1998:HSW, Briane:1998:HTR,
Briane:2002:HNU, Camar:2002:CSD, Cherednichenko:2006:NLH}, materials with 
a macroscopic higher order gradient or non-local elastic response even though they are governed by usual linear
elasticity equations at the microscale \cite{Bouchitte:2002:HSE, Alibert:2003:TMB, Camar:2003:DCS}), 
materials with non-Maxwellian macroscopic electromagnetic behavior \cite{Shin:2007:TDE}, even though they
conform to Maxwell's equations at the microscale, and materials with macroscopic behavior outside that of 
continuum elastodynamics even though they are governed by continuum elastodynamics at the microscale \cite{Milton:2007:NMM}. It is
becoming increasingly apparent that the usual continuum equations of physics do not apply to materials
with exotic microstructures. 

One would really like to be able to characterize the possible macroscopic continuum equations that
govern the behavior of materials, including materials with exotic microstructures. A strategy
for doing this was developed by Camar-Eddine and Seppecher \cite{Camar:2002:CSD, Camar:2003:DCS}). Basically, the idea is to first show that one
can use a continuum construction to model a discrete network, consisting of nodes (terminals) which
are strongly coupled to the continuum matrix and other nodes that are effectively hidden because
they occupy vanishingly small volume and are essentially uncoupled with the continuum matrix [alternatively,
following the ideas of Milton and Willis \cite{Milton:2007:MNS} these nodes might be in an region of the material that is 
declared to be hidden, where the behavior of the fields do not influence the chosen macroscopic
descriptors]. The next step is to characterize the possible responses of discrete networks, in which
one only is interested in the behavior at the terminals. The final step is to characterize the possible
continuum limits of these discrete structures. This program was successfully carried out for
three-dimensional conductivity \cite{Camar:2002:CSD} and three-dimensional linear elasticity \cite{Camar:2003:DCS}, 
giving a complete characterization
of the possible macroscopic equations, under some assumptions such as that the source term does not
vary on the microscale, and that the macroscopic descriptor is a single potential (for electrical
conductivity) or a single displacement field (for linear elasticity).

Our ultimate goal would be to characterize the possible macroscopic electrodynamic, acoustic, and 
elastodynamic equations, achievable under the assumption that the microstructure does 
not vary with time, and also when this assumption is relaxed. A more reachable objective
would be to characterize the macroscopic behavior under the assumption that the fields are
time harmonic, oscillating at a fixed real frequency $\Go$. This paper is devoted to such a characterization,
for discrete dynamical electric networks with grounds, discrete acoustic networks, and discrete elastodynamic networks,
anticipating that this will be key to understanding the possible macroscopic limits in continuum systems.
Curiously, the characterization of the response tensors in the dynamic case turns out to
be easier than in the static case. In the static case, the possible response tensors
of $n$-terminal resistor networks in three-dimensions was essentially characterized by Kirchhoff 
and is known as the generalized $Y-\GD$ theorem: any $n$-terminal network is equivalent
to an $n$-terminal network having no internal nodes and with up to $n(n-1)/2$ resistors
connecting the terminal pairs. 
However, to our knowledge, there is no such 
characterization in two-dimensions. One
exception is for circular planar resistor networks, where the terminals are at the
boundary of a circle, and the network is contained within the circle. For
this class of planar network Curtis, Ingerman, and Morrow \cite{Curtis:1998:CPG} have completely characterized the possible
response matrices. For static $n$-terminal spring networks Camar-Eddine and Seppecher \cite{Camar:2003:DCS} obtained
a complete characterization of the possible response matrices in three-dimensions, but again
the two-dimensional case remains an open problem.

\section{Electrical Circuits}
\setcounter{equation}{0}

\subsection{The lossless electrical case}

To begin with, let us treat the case of an $n$-terminal network consisting only of
capacitors and inductors. An $n$-terminal network is a set of $n+m$ nodes $P_r$. Each pair $(P_r,P_s)$ of nodes may be connected by 
capacitors and/or inductors. The $n$ first nodes, called the terminals of the network, are connected to 
the exterior. When the terminals $P_1$, \dots ,$P_n$ are respectively submitted to voltages
$V_1 e^{-i \omega t}$, \dots ,$V_n e^{-i \omega t}$, the complex currents\footnote{We have chosen to keep track
of the parameters $A_r$ rather than the currents to unify the
mathematics, and make the connections with the discrete elastic models
discussed in this paper more transparent. } entering the $n$ terminals take the form 
$iA_1  e^{-i \omega t}/\Go$, \dots , $i A_n  e^{-i \omega t}/\Go$. 
If we denote, in the same way, $iI_{r,s}e^{-i \omega t}/\Go$ the complex current flowing to node $r$ from node $s$, 
the linear behavior of the capacitors and inductors connecting the two nodes leads to the relation
\beq  I_{r,s}=-I_{s,r}= k_{r,s}(V_s-V_r)\eeq{constitutive}
where the coefficient $k_{r,s}$ is $k_{r,s}=1/L$ for a single inductor while
$k_{r,s}=-\Go^2C$ for a single capacitor, where $L$ is the inductance and $C$
is the capacitance. Of course, $k_{r,s}=0$ when the two nodes are not connected. So any constant $k_{r,s}\in \RE$ is possible.
Note that, in this description different nodes simply joined by wires are considered as a single node. 
We do not allow the terminals to be in this situation (no short circuit). 

When an internal node $P_r$ (with $r>n$) is not connected to the ground,  Kirchhoff's current law must apply. We have
\beq \sum_{s=1}^{n+m} I_{r,s}=0, \eeq{1.1}
(in which we set $I_{r,r}=0$).
At each terminal  $P_r$ (with $r\leq n$) the same law reads
\beq A_r+\sum_{s=1}^{n+m} I_{r,s}=0. \eeq{1.1b}
 
But an internal node can be connected to the ground. At such a grounded node $P_r$, a current can flow toward the ground. 
Equation (\ref{1.1}) does not apply anymore and has to be replaced by
\beq V_r=0. \eeq{1.1g}

We only consider circuits for which $\Go$ is not a resonance frequency.
Then the response $\BA=(A_r)_{r=1}^n$ depends in a linear way on the applied voltages 
 $\BV=(V_r)_{r=1}^n$ : there exists an $n\times n$ matrix $\BW$ with real coefficients $W_{r,s}$ 
such that
\beq A_r=\sum_{s=1}^n W_{r,s} V_s.
\eeq{1.4} 
It is well known that this matrix is symmetric 
\beq W_{r,s}=W_{s,r},\eeq{1.5} 
being the Schur complement of the
matrix characterizing the response when all nodes are regarded as terminals,
which is clearly symmetric.

\medskip Our goal is to characterize the set of matrices $W$ which can be obtained as a response matrix of 
a general (grounded) network but we will also consider two possible restrictions for the networks : 

\medskip\noindent {\bf Ungrounded networks :} In this case, the network is not connected to the ground and, 
at each internal node $P_r$ (with $r>n$), equation (\ref{1.1}) applies. 
In that case the response to a uniform voltage ($V_1=V_2=\dots=V_n$) is zero and the matrix $\BW$ has to satisfy
\beq\forall r,\ \sum_{s=1}^n W_{r,s}=0. \eeq{constr}

\medskip\noindent {\bf Special grounded networks :} For reasons which will become clear in section \ref{acousticsec},
where we treat acoustic networks, 
we pay particular attention to circuits in which inductors are used only to join ungrounded nodes while capacitors 
are only used to join an ungrounded node to a grounded one. Owing to (\ref{1.1g}) the grounded nodes are easily 
eliminated in a first step when computing the response matrix of the circuit and at any node $P_r$ connected 
to a grounded node with a capacitor with capacitance $C_r$, equation (\ref{1.1}) has to be replaced by 
\beq \sum_{s=1}^{n+m} I_{r,s} = -\Go^2 C_r V_r. \eeq{1.1e}
We say that $P_r$ ``has capacitance $C_r$''. Such circuits, we call ``special 
grounded networks'', can then be considered as networks of nodes $P_r$ with capacitance $C_r$ only joined by inductors.

\bigskip
The cases of grounded and ungrounded networks are very similar. Indeed, when considering an ungrounded network, 
we can assume, 
without loss of generality, that one of the terminals, let say $P_n$, has voltage $0$ and decide to call it the ``ground''.
Due to the constraint (\ref{constr}) the response matrix of the network will be determined by 
the $(n-1)\times (n-1)$ {\bf reduced response matrix} $\widetilde \BW=(W_{r,s})_{r,s=1}^{n-1}$ where one
deletes the $n$-th row and column from $\BW$.
Considering $P_n$ as an internal node instead of a terminal, transforms the $n$-terminal ungrounded 
network in a $(n-1)$-terminal grounded one. The response matrix of this network coincides with the reduced matrix of the initial network.
Reciprocally, when considering a grounded network, it suffices to connect all the grounded nodes together, making so a single node, and to consider this node as a new terminal. 
We then obtain an ungrounded new network, the reduced matrix of which corresponds to the response matrix of the initial network.
The problems of finding all possible response matrices for grounded or ungrounded networks are identical as far as there are no topological or physical restrictions 
preventing from connecting together all the grounded nodes.

\bigskip Let us now consider some simple examples of special grounded $n$-terminal networks. 
Of course, the same response matrices may be obtained 
more directly as the response of general grounded networks. 
We also leave the reader to construct the ungrounded networks with the same reduced response matrix.

\begin{Aexa}\label{ex1}
Let $k \in \RE^*$ (the set of non-zero reals)
and consider the simple one-terminal network in which the terminal is connected to an internal node of capacitance 
$C=\frac{k|k|}{(2k-|k|)\Go^2}>0$ with an inductor of inductance $L=\frac {2} {|k|}$ (see fig. \ref{fig:1}).
The response matrix $\BW$ is that of the inductor in series with the capacitance:
\beq \BW=\pmatrix{[L-1/(\Go^2C)]^{-1}}=\pmatrix{k}.\eeq{1.5aa}
\end{Aexa}

Using copies of this circuit with $k>0$ in a network is equivalent 
to allowing the use of internal nodes 
with negative ``capacitance'' when constructing 
special grounded networks.
\begin{figure}
\centerline{\includegraphics{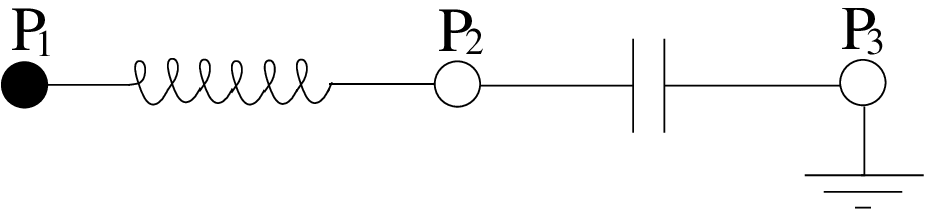}}
\vspace{0.1in}
\caption{Example \ref{ex1}}
\labfig{1}
\end{figure}

\begin{Aexa}\label{ex2}
Let $k \in \RE^*$ and set $C:=\frac{2 |k| + k}{\Go^2}$, $L:=\frac 1 {2|k|}$, $\widetilde C:=\frac{4 (|k| +k)}{\Go^2}$.
We consider the following two-terminal network ($n=2$): terminals 1 and 2 have capacitance $C$. They are joined to an internal node $P_3$ with two inductors of the same inductance $L$. 
$P_3$ has capacitance $\widetilde C$ (see fig. \ref{fig:2}). We let the reader check that the
 response matrix of this circuit is the very elementary matrix
\beq \BW=
\pmatrix{0 & k \cr k & 0}.
\eeq{1.5ac}
\end{Aexa}

\begin{figure}
\centerline{\includegraphics{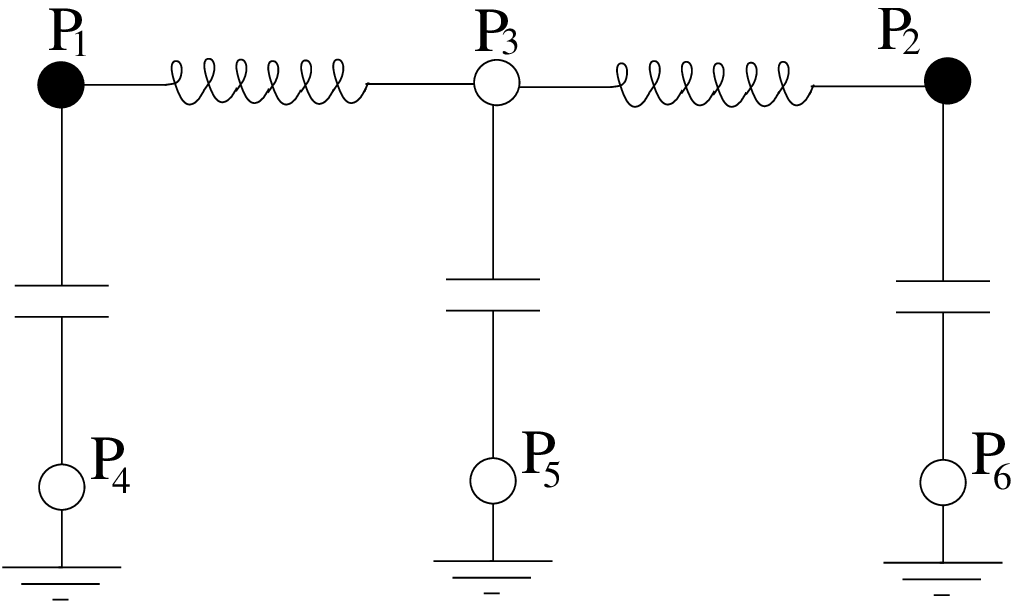}}
\vspace{0.1in}
\caption{Example \ref{ex2} and \ref{ex3}}
\labfig{2}
\end{figure}

\begin{Aexa}\label{ex3}
If we modify only the value of $C$ in the previous example by setting $C:=\frac{2 (|k| + k)}{\Go^2}$, we get the response matrix
\beq \BW=
\pmatrix{-k & k \cr k & -k}.
\eeq{1.5ad}
When $k<0$ this response can more directly be obtained by a unique inductor with inductance $-k^{-1}$ joining the two terminals. 
When $k>0$ the circuit is equivalent to a unique capacitor with capacitance $k \Go^{-2}$ joining the two terminals. 
\end{Aexa}

So, starting with a general grounded network and replacing all capacitors by copies of this circuit leads to a special grounded network 
with the same response matrix. The restriction to special grounded networks does not reduce the set of possible response matrices. 

\medskip \noindent{\bf Superposition principle :}
We assume that all the components (capacitors, inductors, wires and nodes) 
the network is made of occupy arbitrarily small volume. Then, in the three-dimensional case, 
the physical placement of the components has no importance and no crossing problem occurs. 
When considering two networks sharing the same terminals ($P_1$,\dots, $P_n$), by using (if 
necessary) a suitable distortion of one of the networks, we can assume that the internal 
components of the two networks do not intersect. So the response matrix of the network 
obtained by superposition is simply the sum of the response 
matrices of the two initial networks (because the two networks share a common set of voltages
at the terminals, and the current flowing into each terminal is a sum of the currents
flowing into each of the two networks through that terminal).
Note that any $n$-terminal network can be considered 
as a $m$-terminal network (with $m>n$) in 
which $m-n$ terminals are not connected. This superposition property shows also that, in dimension three, 
the problems of finding all possible response matrices for grounded or ungrounded networks are equivalent.

\medskip
Then we easily get the following
\begin{Thm}\label{thm1}
In three-dimensions at a fixed given frequency, any real symmetric 
matrix $\BS$ can be realized as the response matrix of a special grounded network.
It can also be realized as the reduced response matrix of an ungrounded network.
\end{Thm}
\proof Let $n$ be the dimension of the matrix. Let us construct a special grounded network the response matrix of which is $\BS$. 
We first consider the superposition of $n$ copies of Example \ref{ex1}. 
The constant $k$ used in the copy attached to terminal $r$ is chosen by setting $k=S_{r,r}$. 
So the response matrix of the superposition coincides with the diagonal part of $\BS$. 
Then we superimpose on the previous network $\frac {n(n-1)} 2$ copies of Example \ref{ex2}. The constant $k$ used in the copy attached to the pair 
of terminals $(P_r,P_s)$ is chosen by setting  $k=S_{r,s}$. This fixes the off-diagonal elements of the response matrix. 

An ungrounded network, the reduced response matrix of which is $\BS$, can be obtained using the correspondence we already described. 
\qed

\smallskip Note that for ungrounded networks a much simpler construction is possible. Given a real $(n+1)$-dimensional symmetric matrix $\BS$, whose
row sums are zero, we can realize $\BS$ as the response matrix  
of an ungrounded $(n+1)$-terminal network by connecting every pair of terminals $(P_r, P_s)$ 
with a component with constant $k=-S_{r,s}$: then all the off-diagonal elements take their desired values and the diagonal elements automatically 
take the correct values by the constraint (\ref{constr}).

\subsection{The lossy electrical case}

Now let us extend our definition of networks by allowing resistors in the connections between nodes. 
The only change in our analysis is the fact that the constant $k$ in equation (\ref{constitutive}) is no longer real. Indeed for a single resistor 
with resistance $R$ connecting nodes $r$ and $s$ the 
relation (\ref{constitutive}) holds with $k=-i\Go/R$.
More generally $k$ has a negative imaginary part.

We also slightly extend the definition of special grounded networks by allowing any ungrounded node to be joined to a grounded one by a resistive capacitor.
The inductors could also be resistive, but in our constructions we will still require that pairs of ungrounded nodes be connected only by perfect inductors.

The response matrix  $\BW$ of such circuits is complex, symmetric, with negative semidefinite imaginary part,
\beq {\rm Im}\BW\leq  0.\eeq{1.13}
This well-known constraint reflects the second law of thermodynamics that the 
circuit can transform electrical energy into heat but not the reverse. To see this directly it is easy to check
that \eq{1.13} is satisfied for a circuit in which all nodes are terminals, and as a result the quantity
\beq ({\rm Im}\BV)\cdot({\rm Re}\BA)-({\rm Re}\BV)\cdot({\rm Im}\BA)
=-({\rm Re}\BV)\cdot{\rm Im}\BW({\rm Re}\BV)-({\rm Im}\BV)\cdot{\rm Im}\BW({\rm Im}\BV)
\eeq{1.13a}
(which is proportional to the time averaged power dissipation) is always non-negative, where
$\BV=(V_1,V_2,\ldots,V_n)$ and $\BA=(A_1,A_2,\ldots,A_n)=\BW\BV$. This remains true if some of the
$A_r$ are zero which corresponds to a network with internal nodes. Then the left hand side
of \eq{1.13a} is just a sum involving the  $V_r$ and $A_r$ at the terminals and the 
algebraic identity implies \eq{1.13} holds for the response matrix $\BW$ of a network
with internal nodes. It is easy to check that the response matrix has the property \eq{1.13}
if and only if the reduced response matrix has a negative semidefinite imaginary part. 

We have 
\begin{Thm}\label{thm2}
In dimension three, every symmetric complex matrix $\BS$ with negative semi-definite
imaginary part is realizable as the response matrix of some special grounded network. 
It is also realizable as the reduced response matrix of an ungrounded network.
\end{Thm}

\proof Let us first reconsider Example \ref{ex1}. Let $k$ be a complex with negative imaginary part and 
fix  $L=\frac {2} {|k|}$ (which is still a positive real and corresponds to a perfect inductor) and
$C=\frac{k|k|}{(2k-|k|)\Go^2}$ (which has positive real and imaginary parts and then corresponds to a resistive capacitor).
The response matrix is again $\BW=\pmatrix{k}$. Thus we are allowed to use internal nodes 
with any complex ``capacitance'' with positive imaginary part when constructing special grounded networks.

Now, let us consider the real and imaginary parts of  $\BS=\BS^{re}+i \BS^{im}$. They are $n\times n$ symmetric matrices with real coefficients. Thus 
we introduce the $n$ eigenvalues $(k^m)_{m=1}^{n}$ of $\BS^{im}$. 
As $\BS^{im}$ is a negative semidefinite symmetric matrix these eigenvalues  $(k^m)$ are non-positive reals, and the associated
eigenvectors can be chosen to be orthonormal.
We denote by $(a_1^m, a_2^m,\dots a_n^m)$ the $n$-component eigenvector associated with the eigenvalue $k^m$.

Owing to Theorem \ref{thm1} we know that there exists a lossless electrical circuit with $2n$ terminals with the real 
response matrix $\widetilde\BS$ with entries $\widetilde S_{r,s} $ defined by
\beq  \widetilde S_{r,s}= S_{r,s}^{re},\ {\rm if}\ r\leq n\ {\rm and}\ s\leq n,\qquad   \widetilde S_{r,s}=0,\ {\rm if}\ r> n\ {\rm and}\ s> n,\ee
\beq   \widetilde S_{r,n+m}= \widetilde S_{n+m,r}=- k^m\, a^m_r,\ {\rm if}\ r\leq n\ {\rm and }\ m\in\{1,\dots n\}.\ee
Let us endow each terminal $P_{n+m}$ (for $m\in\{1,\dots n\}$) with the 
effective capacity $C_{n+m}=-i {k^m} \Go^{-2}\, $ and consider all these terminals as internal nodes.
At each node  $P_{n+m}$ (for $m\in\{1,\dots n\}$) we have 
\beq k^m  \sum_{s=1}^n a^m_s\, V_s = - \Go^2 C_{n+m} V_{n+m}= i k^m\, V_{n+m}\ee
and, at each terminal $P_r$, for $r\leq n$:
\beq A_r=\sum_{s=1}^{n} S_{r,s}^{re} V_{s} + \sum_{m=1}^{n} - k^m a^m_r\, V_{n+m} . \ee
Using the first equation to eliminate the terms involving $V_{n+m}$ in the second equation, we get
\beq A_r= \sum_{s=1}^n \left( S_{r,s}^{re} +  i \sum_{m=1}^{n} k^m a_r^m a_s^m \right)V_s .\ee
Thus we get the desired response matrix.
\qed

\subsection{The construction in two-dimensions}
If we think of inductors as coils of wires, then it does not make much
physical sense to consider a planar circuit. However metals such as gold or silver or other plasmonic materials can have an electrical
permittivity which is close to being real and negative over certain 
frequency ranges and a rectangular block of such a material can function as an inductor
\cite{Engheta:2005:CEO, Engheta:2007:CLN}.

Now the cases of grounded and ungrounded circuits are quite different. 
In the first case the ground is freely distributed at any internal node 
while in the second case only the nodes which can be connected with a fixed terminal may be grounded.
In two-dimensions we have the important topological restriction
that no two edges are allowed to cross without intersecting at
a common node. This would suggest that the superposition principle
does not apply. Surprisingly, we still have the following

\begin{Thm}\label{thm3}
Every symmetric complex matrix $\BS$ with negative semidefinite
imaginary part is realizable as the response matrix of some planar special grounded network. 
It is also realizable as the reduced matrix of a planar ungrounded network.
\end{Thm}

Note that the circuits described in Examples \ref{ex1}, \ref{ex2} or \ref{ex3} are planar circuits. Let us add a new example.

\begin{Aexa}\label{ex4}
We consider a planar four-terminal ungrounded network. The four terminals are numbered clockwise 1, 2, 3 and 4. Let $k\in \RE^*$. 
We join the pairs of terminals $(P_1,P_2)$, $(P_2,P_3)$, $(P_3,P_4)$ and $(P_4,P_1)$ by four identical components with constant $-k$.  
We introduce an internal node $P_5$ and join each terminal to $P_5$ by components with constant $4k$.
At any terminal $r\in\{1,2,3,4\}$, denoting $\tilde r$ the opposite terminal, equation (\ref{1.1b}) reads
\beq A_r= 4 k (V_5-V_r)-k \sum_{{s=1} \atop {s\not=\tilde r}}^4 (V_s-V_r) \ee
and at node $P_5$, equation (\ref{1.1}) reads
\beq 4k \sum_{s=1}^4 (V_5-V_s)=0.\ee
From these two equations we deduce
\beq  A_r= k\sum_{s=1}^4(V_s-V_r)-k \sum_{{s=1} \atop {s\not=\tilde r}}^4 (V_s-V_r) = k  (V_{\tilde r}-V_r) \ee
The response matrix $\BW$ is
\beq \BW=k \pmatrix{-1&0&1&0\cr 0&-1&0&1\cr 1&0&-1&0\cr 0&1&0&-1}\eeq{1.5f3}
Note that, when $k$ is very large this four-terminal circuit is an approximation of a "virtual crossing". 
The network is equivalent to two connections with constant $k$ joining terminals $1$ and $3$ and terminals $2$ and $4$ without intersecting.

The same  matrix can be obtained as the response matrix of a planar four-terminal special grounded network (under our assumption that the
grounds are allowed to be disconnected from each other).
Indeed it suffices to replace any capacitor by an ad-hoc copy of Example \ref{ex3}.
\end{Aexa}

\begin{figure}
\centerline{\includegraphics{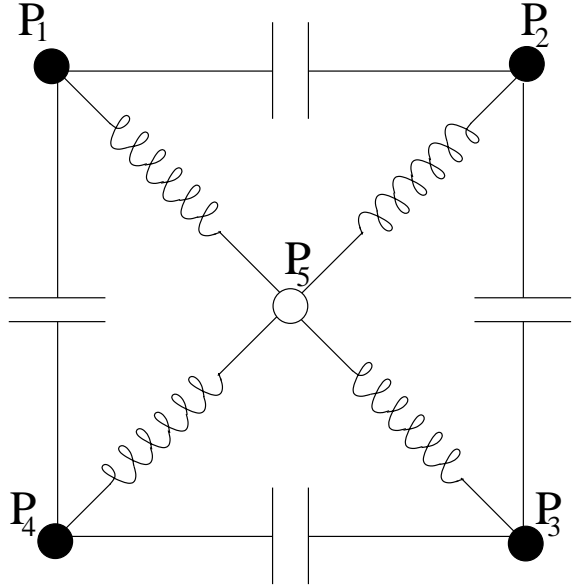}}
\vspace{0.1in}
\caption{Example \ref{ex4}, the virtual crossing when $k>0$. When $k<0$ capacitors and inductors have to be exchanged.}
\labfig{6}

\end{figure}

\bigskip \noindent {\it Proof of theorem \ref{thm3}}. We only consider the case of ungrounded networks. The other case can be treated in a similar way. 
Theorem \ref{thm2} provides a three dimensional ungrounded network which has the desired response matrix. 
A suitable distorsion transforms this circuit in a planar one. 
But the resulting network is not a true planar circuit in the sense that the connections between different pairs of 
nodes $(P_r,P_s)$, $(P_t,P_u)$ cross without any physical interactions. The proof will be completed by proving that any circuit with $p$ crossings is equivalent to another circuit with $p-1$ crossings. 
So a simple induction argument gives us a planar network without any crossing, that is a true planar circuit.

To remove a crossing point, we use a copy of the network described in Example \ref{ex4}.
 Let us isolate a particular crossing of two connections $(P_r,P_s)$, $(P_t,P_u)$, the constants 
of which are denoted respectively $k_{r,s}$ and  $k_{t,u}$. We add four internal 
nodes $P_{r_0}, P_{s_0}, P_{t_0}, P_{u_0} $ and replace the two connections by four 
connections  $(P_r,P_{r_0})$, $(P_s,P_{s_0})$, $(P_t,P_{t_0})$, $(P_u,P_{u_0})$ and a 
copy of Example \ref{ex4} (with an arbitrary constant $k\in\RE^*$ satisfying $k\not=k_{r,s}$ and $k\not=k_{t,u}$  ) as shown in figure \ref{fig:7}.
\begin{figure}
\resizebox{6.0in}{3.0in}{\includegraphics{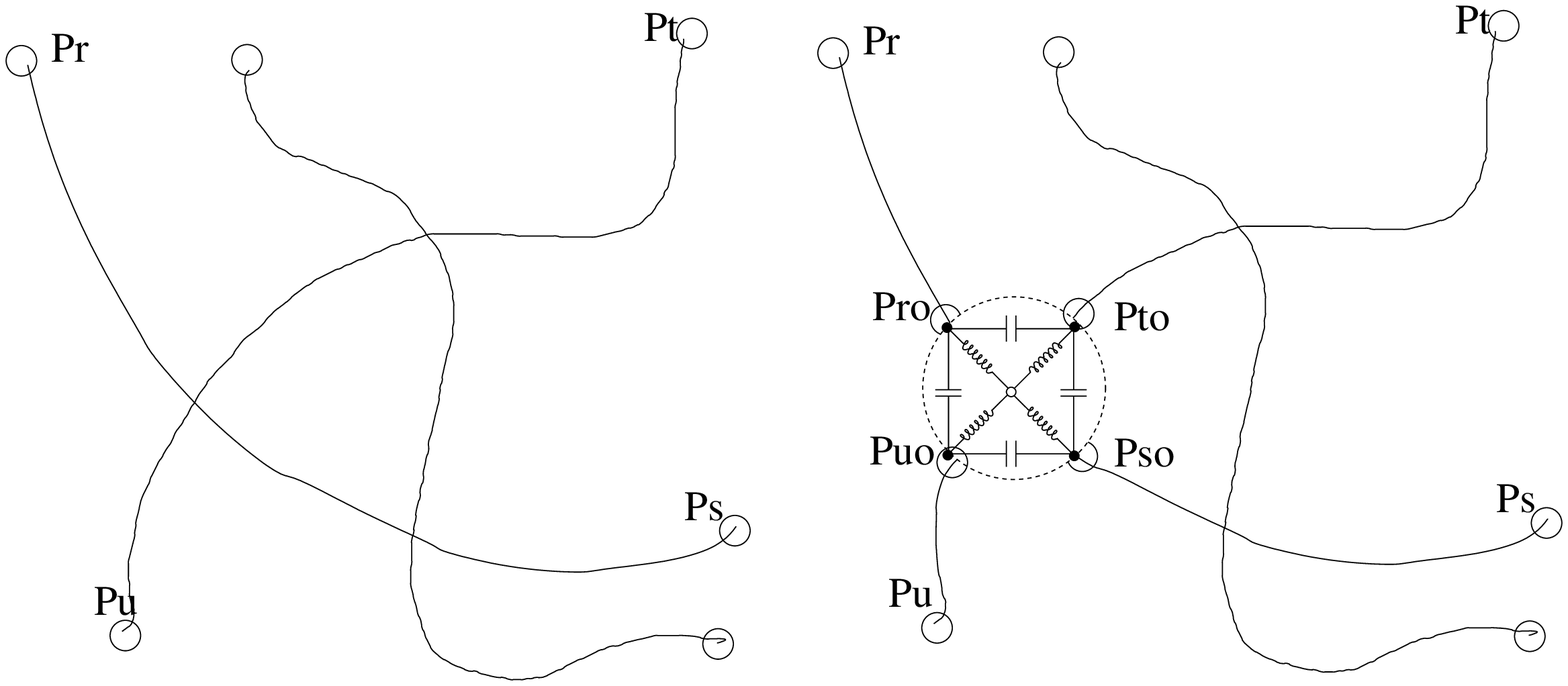}}
\vspace{0.1in}
\caption{The removal of a crossing.}
\labfig{7}
\end{figure}

We choose the constants of the connections  $(P_r,P_{r_0})$, $(P_s,P_{s_0})$, $(P_t,P_{t_0})$, $(P_u,P_{u_0})$ respectively equal to 
\beq k_{r,{r_0}}=2 \left(\frac 1 k_{r,s} - \frac 1 k\right)^{-1}, \quad k_{s,{s_0}}=k_{r,{r_0}} \ee

\beq k_{t,{t_0}}=2 \left(\frac 1 k_{t,u} - \frac 1 k\right)^{-1}, \quad k_{u,{u_0}}=k_{t,{t_0}} \ee
Note that, as $k\in\RE^*$, the imaginary parts of these constants are, like the imaginary parts of  $k_{r,s}$ and $k_{t,u}$, non-positive.
Checking that the response matrix is unchanged is straightforward. \qed

\subsection{Application to the discretized acoustic equation}\label{acousticsec}
The preceeding analysis also applies directly to the discretized acoustic equation.

A domestic water supply network is made of tubes containing a (almost) incompressible fluid and some hydraulic capacitors.
Such a situation can be found also in natural conditions: for instance in an unsaturated porous medium.
Consider a two-dimensional or three-dimensional network of tubes with cavities at the junctions.
Each tube contains a segment of incompressible, non-viscous, fluid with some
density, possibly varying from tube to tube, moving in a time harmonic oscillatory 
manner in response to time harmonic pressures at the junctions. (There could an 
additional time independent constant pressure everywhere, but this does not
affect the equations). 
We define the entire cavity associated with a junction to be the cavity at the junction, plus the remaining 
region in the tubes not occupied by the incompressible fluid.
Each entire cavity contains a compressible, non-viscous, massless fluid with compressibility
possibly varying from junction to junction. The
surfaces between the compressible and incompressible fluids have some surface energy so that the interfaces
remain flat. 

 When the terminals $P_1$, \dots ,$P_n$ are respectively submitted to pressures
$p_1 e^{-i \omega t}$, \dots ,$p_n e^{-i \omega t}$, the complex fluid currents entering the $n$ terminals take the form 
$iA_1e^{-i \omega t}/\Go$, \dots , $i A_n e^{-i \omega t}/\Go$. 
We denote, in the same way, $iI_{r,s}e^{-i \omega t}/\Go$ the complex current flowing to node $r$ from node $s$.
Let $a_{r,s}$ be the cross-sectional area of the tube joining nodes $P_r$ and $P_s$ and let $m_{r,s}$ be the mass of the fluid contained in this tube.
Since the complex force on the fluid segment is $a_{r,s}(p_r-p_s)$ and its complex acceleration is  $I_{r,s}e^{-i \omega t}/a_{r,s}$,
 Newton's law of motion implies
\beq  I_{r,s}=-   I_{s,r}= k_{r,s}(p_s-p_r)\eeq{acoustive}
where $k_{r,s}=a_{r,s}^2/m_{r,s}$. 

Now consider an internal node $P_r$. Due to the motions of the incompressible fluid segments in the tubes the volume
of the associated entire cavity changes with time and the complex pressure $p_r$ in the cavity
adjusts itself according to Hooke's law,
\beq \sum_{s=1}^{n+m} I_{r,s}=-C_r \Go^2 p_r, \eeq{1.19a}
where $C_r=V_r/\Gk$ in which $V_r$ is the volume of the entire cavity when the fluids are 
at rest, and $\Gk$ is the bulk modulus of the fluid in this entire cavity. When the
junction is a terminal the sum must take into account the current entering the terminal. 
If we assume, without loss of generality, that the capacity of each terminal vanishes, we have
\beq A_r+\sum_{s=1}^{n+m} I_{r,s}=0. \eeq{1.19b}

Since \eq{acoustive}, \eq{1.19a} and \eq{1.19b} are the direct analogs of equations \eq{constitutive} \eq{1.1e} and 
\eq{1.1b} which describe special grounded networks, all the previous analysis and associated theorems apply. 
In particular, to get a negative
effective ``bulk modulus'' in a cavity we just follow the approach of Fang et.al.\cite{Fang:2006:UMN}
and connect that cavity to a Helmholtz resonator, i.e. connect it 
to another cavity containing compressible fluid with a tube containing a plug of 
incompressible fluid with mass chosen so that the system is above resonance.

The lossy case arises when the
fluid in some, or all, of the junctions has some bulk viscosity,
so that $\Gk$ has some negative imaginary part, and consequently
$C_r$ in equation \eq{1.19a} has a positive imaginary part. The lossy case
also arises if the incompressible fluid segments have some shear viscosity
so that Darcy's law implies $k_{r,s}$ has a complex part due to the fluid permeability of the 
tube.

We can conclude that any response is possible for an acoustic discrete system 
provided that the total dissipation is non-negative.

\begin{figure}
\vspace{2in}
\hspace{1.0in}
{\resizebox{2.0in}{1.0in}
{\includegraphics[0in,0in][6in,3in]{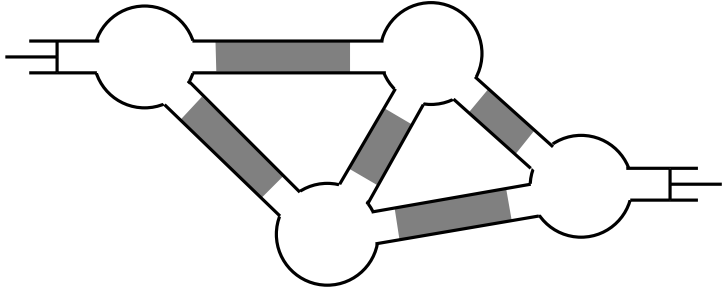}}}
\vspace{0.1in}
\caption{A two-terminal discrete acoustic network. In the idealized model the four cavities 
contain compressible massless fluid, while the grey shaded fluid plugs in the five tubes 
contain incompessible fluid with some mass. The response of the network is measured
by the movement of the two frictionless pistons, in response to the time harmonic forces
acting on them which control the pressures in the terminal cavities.}
\labfig{acoust}
\end{figure}

\section{Discrete Elastodynamics}
\setcounter{equation}{0}

We now turn our attention to networks of springs with point masses at the nodes.
We emphasize that our analysis is for idealized linear networks and that
we do not consider questions of stability, and in particular stability
against buckling. 

Let us denote $\Bu_r e^{-i \Go t}$ the displacement of node $P_r$ ($\Bu$ is a three dimensional complex vector)
and $\BF_{r,s}e^{-i \Go t}$ the force exerted on node $P_r$ by the spring (if any) joining $P_s$ and $P_r$.
In the same way, for any terminal $P_r$ ($r\leq n$), 
let us denote $\BA_{r} e^{-i \Go t}$ the additional external force applied on the system at that terminal.
At each interior node $P_r$, Newton's law applies and we have 
\beq \sum_{s=1}^{n+m}\BF_{r,s}=-m_r\Go^2\Bu_r, \eeq{2.1}
(which is the analog of \eq{1.1e})
where $m_r$ denotes the mass of node $P_r$. At a terminal $P_r$ ($r\leq n$), we have also to take 
into account the external force:
\beq \BA_r+\sum_{s=1}^{n+m}\BF_{r,s}=-m_r\Go^2\Bu_r. \eeq{2.1b}

\subsection{The purely elastic case}
To begin with, let us treat the case of purely elastic $n$-terminal networks 
where there is no damping in the springs.
Between each pair of nodes $P_r$ and $P_s$, located at positions $\Bx_r$ and $\Bx_s$ 
that are linked by a spring, Hooke's law applies, and we have 
\beq \BF_{r,s}=-\BF_{s,r}=k_{r,s}\Bn_{r,s}\otimes\Bn_{r,s}\cdot(\Bu_s-\Bu_r),
\quad {\rm where}~\Bn_{r,s}=\frac {\Bx_s-\Bx_r}{\|\Bx_s-\Bx_r\|},\eeq{2.2}
(which is the analog of \eq{constitutive})
where $k_{r,s}=k_{s,r}$ is the (positive real) spring constant.

The elastodynamic response of the network is governed by a matrix $\CW$
which has second order tensors $\BW_{r,s}$ as its entries, and links the set
of forces $\CA=(\BA_1,\BA_2,\ldots,\BA_n)$ with
the set of displacements $\CU=(\Bu_1,\Bu_2,\ldots,\Bu_n)$, through
the relation
\beq \BA_r=\sum_{s=1}^n\BW_{rs}\Bu_s.
\eeq{2.3} 
This matrix $\CW$ is real and has the symmetry property that, for any $r$, $s$ in $\{1,\dots,n\}$,
\beq \BW_{r,s}=(\BW_{s,r})^T.
\eeq{2.4} 
and the proof of this property is similar to the proof of \eq{1.5} in the electrical case.

\begin{Aexa}\label{ex1b}
The simplest non-trivial two terminal network just consists of terminals $P_1$ and $P_2$ joined by
a spring with constant $k$ and direction $\Bn=\Bn_{1,2}$. According to \eq{2.2} the matrix $\CW$ is
\beq \CW=\pmatrix{k\,\Bn\otimes\Bn & -k\,\Bn\otimes\Bn \cr
                  -k\,\Bn\otimes\Bn & k\,\Bn\otimes\Bn}.
\eeq{2.4a}
\end{Aexa}

\begin{Aexa}\label{ex2b}
Let us consider the two terminal network in which the terminals $P_1$ and $P_2$ are joined to three 
internal nodes $P_3$, $P_4$, $P_5$ making two non-degenerate simplexes $(P_1,P_3,P_4,P_5)$ and $(P_2,P_3,P_4,P_5)$. 
To each edge of these simplexes corresponds a spring. Nodes have no mass.
In structural mechanics such a structure is called a simple truss. Its response vanishes when 
the applied displacements $\Bu_1$, $\Bu_2$ correspond to a rigid motion. So the response matrix 
corresponds to a non-negative quadratic form depending only on $(\Bu_1-\Bu_2)\cdot \Bn_{1,2}$. It takes the form (\ref{2.4a}). 
The constant $k$ can be tuned by multiplying all the constants of the truss by a common positive factor.
\end{Aexa}

Note that the choice of the position of the internal nodes is quite free. 
A given finite set of points can easily be avoided. Moreover, in dimension three, the internal nodes can be 
chosen in such a way that the five segments $(P_1,P_3)$, $(P_1,P_4)$ $(P_2,P_3)$, $(P_2,P_4)$ $(P_3,P_4)$ 
do not intersect a given finite set of straight lines (but maybe at terminals $P_1$, $P_2$).

\begin{Arem}\label{strrep}
Replacing a single spring in a network by a structure described in Example \ref{ex2b} will not change the response matrix of the network. 
In dimension three, making all the needed replacements we can restrict our attention to networks in which any different springs 
do not intersect and have different directions.
\end {Arem}

\begin{Aexa}\label{ex3b}
Let $\mu$ be a non vanishing real and $\Bn$ a unit vector. Consider the very simple one-terminal spring network, where there is only 
one spring with constant $k=k_{1,2}=\frac {\mu |\mu|}{2\mu-|\mu|} \Go^2$ linking terminal $P_1$ with a single interior node $P_2$ chosen in such a 
way that $\Bn_{1,2}=\Bn$. Terminal $P_1$ has no mass while the mass of node $P_2$ is $m=\frac {|\mu|} 2$. 
We have the equations
\beq  \BF_{2,1}= k\, \Bn\otimes\Bn\cdot(\Bu_1-\Bu_2) =-m \Go^2\, \Bu_2= \BA_1.\ee
The elimination of $\Bu_2$ leads to $\BA_1=-\frac {k m \Go^2} {k-m\Go^2}\, \Bn\otimes\Bn \cdot \Bu_1  = -\mu\Go^2\Bn\otimes\Bn \cdot \Bu_1$. 
This system endows $P_1$ with the  tensor valued ``effective mass'' :
$\BM=\mu \Bn\otimes \Bn$ which can either
be positive or negative semidefinite depending on the sign of $\mu$. The physical reason that one can obtain negative
values of $\mu$ is that these are achieved when the spring-mass system is above resonance, i.e. when $k-m\Go^2<0$, and
as a result the mass oscillates $180^\circ$ out of phase with the motion of the terminal. 
\end{Aexa}

Note that the choice of the position of the internal node is free on the straight line $(\Bx_1,\Bn)$. 
Thus a given finite set of points can easily be avoided. The spring can also be replaced using remark \ref{strrep}
in order to avoid intersections with a given finite set of straight lines. 

Now let $\BM$ be any real symmetric tensor. 
Superimposing up to three copies of the previous structure choosing for $\mu$ the eigenvalues of $\BM$ and 
for $\Bn$ the corresponding eigenvectors of $\BM$ we get 

\begin{Arem}\label{effmass}
Any node can be endowed by any real symmetric tensor effective mass.
\end {Arem}

\begin{Aexa}\label{ex4b}
Let $K$ be a real, $\Bn_1$, $\Bn_2$ be two unit vectors and $\Bx_1$, $\Bx_2$ be two distinct points 
such that at least one of the two vectors $\Bn_1$, $\Bn_2$ is not in the direction $\Bx_2-\Bx_1$. We
consider the two terminal network consisting of terminals $P_1$, $P_2$ at points  $\Bx_1$, $\Bx_2$ and two internal nodes $P_3$, $P_4$ 
placed in such a way that $\Bn_{1,3}=\Bn_1$, $\Bn_{2,4}=\Bn_2$ and $\Bv:=\Bn_{3,4}$ is neither colinear with $\Bn_1$ nor $\Bn_2$.
We introduce two new unit vectors  $\Bw_1$ and $\Bw_2$ which complete respectively the basis $(\Bn_1, \Bv)$ and  $(\Bn_2, \Bv)$.

Springs with constant $k=|K|$ join pairs $(P_1,P_3)$ and $(P_2,P_4)$. A spring with constant $k'=2|K|-K$ joins $(P_3,P_4)$.
Nodes $P_3$ and $P_4$ are endowed respectively with the effective masses 
$k'\Go^{-2}  (\Bv\otimes \Bn_1  + \Bn_1\otimes\Bv + \Bw_1\otimes \Bw_1)$ 
and $k'\Go^{-2} (\Bv\otimes \Bn_2  + \Bn_2\otimes\Bv + \Bw_2\otimes \Bw_2)$.

As noticed in Example \ref{ex3b} such effective masses need the introduction of extra springs and internal nodes. 
Again we have a large freedom in the choice of the position of the nodes $P_3$, $P_4$ on the lines $(\Bx_1,\Bn_1)$ and $(\Bx_2,\Bn_2)$ 
and we can avoid any given finite set of points. Owing to Remark \ref{strrep}
we can also construct this structure avoiding any intersection with a given finite set of straight lines.

Let us introduce the dual basis $(\Bn_1^*, \Bv^* ,\Bw_1^*)$ of $(\Bn_1, \Bv ,\Bw_1)$ (i.e. satisfying $\Bn_1^*\cdot\Bn_1=1$, 
$\Bn_1^*\cdot\Bv=0$, $\Bn_1^*\cdot\Bw_1=0$, $\Bv^*\cdot\Bn_1=0$, $\Bv^*\cdot\Bv=1$, $\Bv^*\cdot\Bw_1=0$,  
$\Bw_1^*\cdot\Bn_1=0$,  $\Bw_1^*\cdot\Bv=0$ and $\Bw_1^*\cdot\Bw_1=1$) and in the same way the dual basis  $(\Bn_2^\circ, \Bv^\circ, \Bw_2^\circ)$ of $(\Bn_2, \Bv, \Bw_2)$.
The displacement of nodes $P_3$, $P_4$ are respectively written in the form 
\beq \Bu_3=a  \Bn_1^* + b  \Bv^* +c \Bw_1^*,\qquad \Bu_4=d \Bn_2^\circ +e \Bv^\circ+f \Bw_2^\circ. \ee
At nodes  $P_3$ and $P_4$, equation (\ref{2.1}) reads
\begin{eqnarray}
-k (\Bn_1\cdot \Bu_1)\Bn_1 + k a \Bn_1 - k' (e-b)\Bv &=& k' (a \Bv + b \Bn_1 + c \Bw_1) \cr
-k (\Bn_2\cdot \Bu_2)\Bn_2 + k d \Bn_2 + k' (e-b)\Bv &=& k' (d \Bv + e \Bn_2 + f \Bw_2)
\end{eqnarray}
from which we deduce $b-e=a=-d= \frac K {|K|} (\Bn_1\cdot \Bu_1-\Bn_2\cdot \Bu_2) $ and $c=f=0$. 
Now at terminals $P_1$ and $P_2$  equation (\ref{2.1b}) reads
\begin{eqnarray}
\BA_1&=&k (\Bn_1\cdot \Bu_1)\Bn_1 - k a \Bn_1 - \Go^2 \BM_1\cdot \Bu_1 \cr
\BA_2&=&k (\Bn_2\cdot \Bu_2)\Bn_2 - k d \Bn_2 - \Go^2 \BM_2\cdot \Bu_2,
\end{eqnarray}
where $\BM_1$ and $\BM_2$ denote the effective masses of terminals $P_1$, $P_2$ which we have not yet fixed.
Then the response matrix of the network is then
\beq \CW=\pmatrix{
- \Go^2 \BM_1+ (|K|-K)\Bn_1\otimes \Bn_1 & K \Bn_1\otimes \Bn_2 \cr
K \Bn_2\otimes \Bn_1  & - \Go^2 \BM_2+ (|K|-K) \Bn_2\otimes \Bn_2}\ee
The key feature of this response matrix is that the off-diagonal matrix is
proportional to $\Bn_1\otimes \Bn_2$. This could have been anticipated since
the spring joining terminals $(P_1,P_3)$ exerts a force on $P_1$ in the direction $\Bn_1$,
and this force can only depend on $\Bu_2$ through the component of $\Bu_2$ in the direction $\Bn_2$
of the spring joining terminals $(P_2,P_4)$.

\end{Aexa}

\begin{Aexa}\label{ex4ba}
Considering in the previous example the particular case $\Bn_1=\Bn_2=\Bn$ 
and  choosing the appropriate values for the tensors $\BM_1$ and $\BM_2$ we get
\beq \CW=\pmatrix{
 -K \Bn\otimes \Bn  & K \Bn\otimes \Bn \cr
K \Bn\otimes \Bn  &  -K \Bn\otimes \Bn},  \eeq{strangeressort}
which is similar to the response matrix of a single spring 
but where the constant $K$ can be negative. More important
is the fact that in this structure the direction of action $\Bn$ 
is no longer correlated with the direction of the vector $\Bx_2-\Bx_1$. 

Remember however that we have the restriction that $\Bn$ cannot be in the direction $\Bx_2-\Bx_1$. 
But we can get rid of this restriction by considering two copies of this structure: 
one of these copies joins terminal $P_1$ to an internal node $P_3$ placed at a point $\Bx_3$ 
such that  $\Bn_{1,3}$ is not parallel to $\Bx_2-\Bx_1$ while the other one joins terminal $P_2$ to $P_3$. 
In both copies the constant is $2K$ and the direction of action is $\Bn=\Bn_{1,2}$. 
It is easy to check that the response matrix of such a structure is still given by (\ref{strangeressort}). 
In that way we actually get a virtual spring with possibly negative spring constant.

\end{Aexa}

This makes free the position of the internal nodes: 
indeed, in any network, we can change the position $\Bx_r$ of a node $P_r$ to any other position $\Bx_r'$, 
replacing all the springs joining $P_r$ to other nodes $P_s$ 
by a copy of Example \ref{ex4ba} with $\Bn_1=\Bn_{r,s}$. 
Clearly the response matrix will remain unchanged. Thus we have 

\begin{Arem}\label{ndrep}
Any network has an equivalent network the internal nodes of which avoid a given finite set of points.
\end {Arem}

Combining Remark \ref{ndrep} with Remark \ref{strrep} enables us to assume when considering two different networks that they 
do not share any internal node and that the springs of the different networks do not intersect. Then we have 

\begin{Arem}\label{super}  Superposition principle : In three dimensions, if $\CW_1$ and $\CW_2$ are two
realizable response matrices each associated with $n$-terminals
in the same positions $P_1, P_2, \ldots, P_n$, then the response
matrix $\CW_1+\CW_2$ is also realizable. The network which realizes
the matrix $\CW_1+\CW_2$ is just a superposition of suitable modifications of the networks which realize $\CW_1$ and $\CW_2$.
\end {Arem}

\begin{Aexa} \label{ex4bb} Choosing in Example \ref{ex4b} the appropriate values for the tensors $\BM_1$ or $\BM_2$ we can get
\beq \CW=\pmatrix{
0 & K \Bn_1\otimes \Bn_2 \cr
K \Bn_2\otimes \Bn_1  &  0 } \ee
\end{Aexa}

As any matrix $\BW$ is the sum of rank one matrices, 
the superposition principle implies that, for any  matrix $\BW$, there exists a two-terminal network the response matrix of which is
\beq \CW=\pmatrix{
0 & \BW \cr
\BW^T  &  0 }. \ee
And we obtain the following

\begin{Thm}\label{thm4}
In three dimensions given any set of $n$ points $\Bx_1,\Bx_2, \ldots \Bx_n$, and any
real $n\times n$ matrix $\CW$ with second order tensor entries $\BW_{ij}$
satisfying the symmetry properties \eq{2.4},
then there is purely elastic network with terminals $P_1,P_2, \ldots P_n$ at positions  $\Bx_1,\Bx_2, \ldots \Bx_n$ and
realizing $\CW$ as its response matrix.
\end{Thm}

\proof  It is enough  to attach to each pair $(P_r,P_s)$ a copy of Example \ref{ex4bb} in which $\BW$ is chosen to be $\BW_{r,s}$ , 
then to endow each terminal $P_r$ with the effective mass corresponding to the symmetric matrix $\BW_{r,r}$. Then we conclude using the superposition principle.\qed

\subsection{Elastodynamic networks with damping}

An elastodynamic network with damping is a network with point masses at the nodes and viscoelastic springs joining the nodes.
If we allow viscous damping in the springs, the constant $k$ in (\ref{2.2}) becomes complex with a non-positive imaginary part.
(The real part of $k$ is still non-negative, and masses are still non-negative reals.)
Then the matrix $\CW$ is 
complex and symmetric, with negative semidefinite imaginary part, 
\beq {\rm Imag}\CW\leq 0,
\eeq{2.22}
which reflects the second law of thermodynamics that averaged over time the 
network can transform mechanical energy into heat, but not the reverse. The proof
of \eq{2.22} is similar to the electrical case.

Let us revisit the examples we gave in the previous section. Examples \ref{ex1b} and \ref{ex2b} are unchanged: the constant $k$ 
in the response matrix is now complex with a positive real part and a negative imaginary part. 
Example \ref{ex3b} is still valid : indeed for any complex $\mu$ with positive imaginary part the constant $k$ defined by
$k=\frac {\mu |\mu|}{2\mu-|\mu|} \Go^2$  has a negative imaginary part and a positive real part. Then remark \ref{effmass} can be generalized in

\begin{Arem}\label{effmasscomp}
Any node can be endowed by any complex symmetric tensor effective mass provided that its imaginary part is positive semidefinite.
\end {Arem}

\begin{Thm}\label{thm5}
In three dimensions, given any set of $n$ points $\Bx_1,\Bx_2, \ldots \Bx_n$, and any
complex matrix $\CW$ with second order tensor entries $\BW_{r,s}$
satisfying the symmetry properties \eq{2.4}
and the constraint \eq{2.22},
then there is an elastodynamic network with damping realizing $\CW$ as its response matrix,
\end{Thm}

\proof Let us consider the real and imaginary parts of  $\CW=\CW^{re}+i \CW^{im}$. They are $n\times n$ matrices with $3\times 3$ real 
entries denoted respectively $\BW_{r,s}^{re}$ and   $\BW_{r,s}^{im}$ and can be identified with $3n\times 3n$ symmetric matrices with real coefficients.
Thus we introduce the $3n$ eigenvalues $(k^m)_{m=1}^{3n}$ of $\CW^{im}$. 
As $\CW^{im}$ is a negative semidefinite symmetric matrix these eigenvalues  $(k^m)$ are non-positive reals, and
the corresponding eigenvectors can be chosen to be an orthonormal set. 
We denote by $(\Ba_1^m, \Ba_2^m,\dots \Ba_n^m)$ a $3n$-component eigenvector (identified with an $n$-entry vector where the entries are 
$3$-component vectors) associated with the eigenvalue $k^m$ and we introduce an extra unit vector $\Bb$.

Owing to Theorem \ref{thm4} we know that there exists a (no damping) elastic network with $4n$ terminals with the real response matrix $\widetilde\CW$ with entries 
$\widetilde\BW_{r,s} $ defined by
\beq  \widetilde\BW_{r,s}=\BW_{r,s}^{re},\ {\rm if}\ r\leq n\ {\rm and}\ s\leq n,\qquad   \widetilde\BW_{r,s}=0,\ {\rm if}\ r> n\ {\rm and}\ s> n,\ee
\beq   \widetilde\BW_{r,n+m}= \widetilde\BW_{n+m,r}^T=- k^m\, \Ba_r^m \otimes \Bb,\ {\rm if}\ r\leq n\ {\rm and }\ m\in\{1,\dots 3n\}.\ee

Then, owing to Remark \ref{effmasscomp}, let us now endow each terminal $P_{n+m}$ (for $m\in\{1,\dots 3n\}$) with the 
effective mass tensor $\BM_{n+m}=-i{k^m} \Go^{-2}\, \BI$ and consider all these terminals as internal nodes.
At each node  $P_{n+m}$ (for $m\in\{1,\dots 3n\}$) we have 
\beq k^m  \sum_{r=1}^n (\Ba_r^m\cdot \Bu_r)\, \Bb = -\Go^2 \BM_{n+m} \cdot \Bu_{n+m}= i k^m \Bu_{n+m}\ee
and, at each terminal $P_r$, for $r\leq n$:
\beq \BA_r=\sum_{s=1}^{n} \BW_{r,s}^{re} \cdot \Bu_{s} + \sum_{m=1}^{3n} - k^m  (\Bb\cdot \Bu_{n+m}) \Ba_r^m . \ee
Using the first equation to eliminate the terms involving $\Bu_{n+m}$ in the second equation, we get
\beq \BA_r= \sum_{s=1}^n \left(\BW_{r,s}^{re} +  i \sum_{m=1}^{3n} \left( k^m (\Ba_r^m\otimes \Ba_s^m)\right) \right) \cdot \Bu_s .\ee
Thus we get the desired response matrix.
\qed

\subsection{Planar elastodynamic networks}
As in the electrical case, in two-dimensions, we have the important topological restriction
that no two edges are allowed to cross without intersecting at
a common node. Now we have the additional restriction that a spring
between two nodes must lie along the segment joining those two nodes. 
Despite these restrictions we have:

\begin{Thm}
\label{thm6}
Theorems \ref{thm4} and \ref{thm5} still hold true for planar networks.
\end{Thm}
\proof 
Example \ref{ex2b} can be adapted to the planar case : the simplexes we used are now simply triangles. However Remark \ref{strrep} is no longer valid. 
Due to the topological restrictions, Example \ref{ex2b} cannot be used to avoid crossings. It still can be used to change the 
direction of the springs in a network. So we can assume that any crossing point is a generic one, which means that 
only two springs are crossing at that point and the angle they make is non-zero.

Let us allow, for a while, springs to intersect without interacting and let us call {\it pseudo-planar} such networks.
In this setting, Example \ref{ex3b} and \ref{ex4b} and the superposition principle are still valid. Nothing is changed from the 
three dimensional case : we can construct a pseudo-planar network with any desired response matrix. Owing to the previous 
remark we can assume that all crossing points in this  pseudo planar network are generic ones.

Now let us consider two crossing springs (let us say connecting $(P_1,P_2)$ and connecting $(P_3,P_4)$ with constants $k_{1,2}$ and  $k_{3,4}$) in this network and let us replace these two springs by the following network:

\begin{Aexa}\label{ex5b} Let $k_{1,2}$ and $k_{3,4}$ be any positive reals 
(or complex with positive real part and negative imaginary part) and consider the 4-terminal network 
where the terminals $P_i$ ($i\leq 4$) are placed at points $\Bx_i$ such that the segments $[\Bx_1,\Bx_2]$ and $[\Bx_3,\Bx_4]$ 
have an intersection at a single point $\Bx_5$. The network has an internal node $P_5$ at point $\Bx_5$.
We assume that the nodes have no mass and that four springs join  $P_1$, $P_2$, $P_3$, $P_4$ to $P_5$ with constants respectively equal to 
$k_{1,5}=k_{2,5}:=2 k_{1,2}$  and $k_{3,5}=k_{4,5}:=2 k_{3,4}$. We have
\beq \BA_1=\BF_{1,5}=- 2 k_{1,2}(\Bn_{1,2}\otimes\Bn_{1,2})\cdot(\Bu_1-\Bu_5),\quad
\BA_2= \BF_{2,5}=- 2 k_{1,2}(\Bn_{1,2}\otimes\Bn_{1,2})\cdot(\Bu_2-\Bu_5), \ee
\beq \BA_3=\BF_{3,5}=-2 k_{3,4}(\Bn_{3,4}\otimes\Bn_{3,4})\cdot(\Bu_3-\Bu_5),\quad
\BA_4=\BF_{4,5}=- 2 k_{3,4}(\Bn_{3,4}\otimes\Bn_{3,4})\cdot(\Bu_4-\Bu_5),\ee
\beq \BF_{1,5}+\BF_{2,5}+\BF_{3,5}+\BF_{4,5}=0.\ee
Owing to the geometrical assumptions $(\Bn_{1,2}, \Bn_{3,4})$ makes a basis and we introduce its dual basis $(\Bn_{1,2}^*, \Bn_{3,4}^*)$. 
Writing $\Bu_5=a \Bn_{1,2}^* + b \Bn_{3,4}^*$, the previous system of equations becomes
\beq \BA_1=-2 k_{1,2}( \Bn_{1,2}\cdot \Bu_1- a ) \Bn_{1,2}, \quad
\BA_2=- 2 k_{1,2}(\Bn_{1,2}\cdot \Bu_2- a ) \Bn_{1,2}\ee
\beq \BA_3=-2 k_{3,4}(\Bn_{3,4}\cdot  \Bu_3- b ) \Bn_{3,4}, \quad
\BA_4=- 2 k_{3,4}(\Bn_{3,4}\cdot \Bu_4 -b ) \Bn_{3,4}\ee
\beq  \Bn_{1,2}\cdot ( \Bu_1 +  \Bu_2)= 2a, \quad \Bn_{3,4}\cdot  (\Bu_3 + \Bu_4) = 2 b  \ee
The elimination of $\Bu_5$ (i.e. of $a$ and $b$) in this system leads to 
\beq  \BA_1=-\BA_2=k_{1,2}(\Bn_{1,2}\otimes\Bn_{1,2})\cdot(\Bu_1-\Bu_2),\qquad
 \BA_3=-\BA_4=k_{3,4}(\Bn_{3,4}\otimes\Bn_{3,4})\cdot(\Bu_3-\Bu_4)\ee
The response matrix of this network is equivalent to the response of two springs joining directly and independently  $P_1$ to $P_2$ and 
$P_3$ to $P_4$ with constants $k_{1,2}$ and $k_{3,4}$.
\end{Aexa}

\medskip Replacing two crossing springs by a copy of Example \ref{ex5b} removes a crossing point (and does not create any new one). 
Hence we can successively remove all crossing points in the pseudo planar network and obtain a true planar 
network with the desired response matrix. 
This analysis is valid in both purely elastic and damping cases. \qed

\section*{Acknowledgements}
Graeme Milton is grateful for support from 
the Universit{\'e} de Toulon et du Var and from the
National Science Foundation through grant DMS-070978.The authors are
grateful to Fernando Vasquez for comments on the manuscript.

\bibliography{/u/ma/milton/tcbook,/u/ma/milton/newref}

\end{document}